\documentclass[prl,twocolumn,showpacs,preprintnumbers,amsmath,amssymb,superscriptaddress]{revtex4}

\usepackage{graphicx}
\usepackage{epsfig}
\usepackage{amsmath}
\usepackage{amssymb}
\usepackage{textcomp}
\usepackage{dcolumn}
\usepackage{bm}
\usepackage{braket}
\usepackage[usenames,dvipsnames]{xcolor}
\usepackage[colorlinks=true,citecolor=MidnightBlue,linkcolor=MidnightBlue,urlcolor=MidnightBlue]{hyperref}

\PassOptionsToPackage{numbers,sort&compress}{natbib}

\makeatletter
\g@addto@macro\normalsize{%
  \setlength\abovedisplayskip{4pt}
  \setlength\belowdisplayskip{4pt}
  \setlength\abovedisplayshortskip{4pt}
  \setlength\belowdisplayshortskip{4pt}
}
\makeatother

\newcommand{\BibitemShut}[1]{} 

\begin{document}

\newcommand*{\MAINZ}{QUANTUM, Institut f\"ur Physik, Johannes Gutenberg-Universit\"at Mainz, Staudingerweg 7, 55128 Mainz, Germany}
\affiliation{\MAINZ}
\newcommand*{\ERLANGEN}{Institut f\"ur Optik, Information und Photonik, Friedrich-Alexander Universit\"at Erlangen-N\"urnberg, Staudtstr. 1, 91058 Erlangen, Germany}
\newcommand*{\SAOT}{Erlangen Graduate School in Advanced Optical Technologies (SAOT), Friedrich-Alexander Universit\"at Erlangen-N\"urnberg, Paul-Gordan-Str. 6, 91052 Erlangen, Germany}
\homepage{http://www.quantenbit.de}

\title{Light from an ion crystal: bunching or antibunching?}
\author{Sebastian Wolf}\email{wolfs@uni-mainz.de}\affiliation{\MAINZ}
\author{Stefan Richter}\affiliation{\ERLANGEN}\affiliation{\SAOT}
\author{Joachim von Zanthier}\affiliation{\ERLANGEN}\affiliation{\SAOT}
\author{Ferdinand Schmidt-Kaler}\affiliation{\MAINZ}

\date{\today}

\begin{abstract}
Photon statistics divides light sources into three different categories, characterized by bunched, antibunched or uncorrelated photon arrival times. Single atoms, ions, molecules, or solid state emitters display antibunching of photons, while classical thermal sources exhibit photon bunching. Here we demonstrate a light source in free space, where the photon statistics depends on the direction of observation, undergoing a continuous crossover between photon bunching and antibunching. We employ two trapped ions, observe their fluorescence under continuous laser light excitation, and record the spatially resolved autocorrelation function $g^{(2)}(\tau)$ with a movable Hanbury Brown and Twiss detector. Varying the detector position we find a minimum value for antibunching, $g^{(2)}(0) = 0.60(5)$ and a maximum of $g^{(2)}(0)=1.46(8)$ for bunching, demonstrating that this source radiates fundamentally different types of light alike. The observed variation of the autocorrelation function is understood in the Dicke model of heralded entangled states and the observed maximum and minimum values are modeled, taking independently measured experimental parameters into account.   
 
\end{abstract}


\maketitle
Hanbury Brown and Twiss were the first to measure the arrival times of photons emitted by classical sources, i.e., from arc lamps and stars, and found that the autocorrelation function $g^{(2)}(\tau)$ for thermal light takes a value of 2 for zero time delay, thus observing photon bunching \cite{brown1956correlation,brown1956test}. By contrast, for a single-photon source like a single atom, ion, molecule, or solid state emitter antibunching is observed with $g^{(2)}(0)=0$, as after the release of a photon the re-excitation of the emitter takes time  \cite{loudon2000quantum}.  This was observed from single atoms, ions and solid state emitters \cite{kimble1977photon,diedrich1987nonclassical,basche1992photon,aharonovich2016solid}. Arrays of quantum emitters, however, are predicted to show even more fascinating collective free-space radiation properties resulting in emitter spatial distributions with either superradiant or subradiant states \cite{guimond2019subradiant}. Only if the spatial radiation properties of the ensemble are averaged out, e.g., by collecting a large solid angle into the detector, those features wash out and as a reminder the contrast of superradiance and subradiance is reduced.
Yet the spatial dependence of $g^{(2)}(\tau)$ for two identical indistinguishable and immobile single-photon emitters in free space has not been measured so far - what autocorrelation function will be detected? 

It turns out that the $g^{(2)}$ function depends on the observation and may vary between bunched (i.e., super-poissonian), uncorrelated (poissonian) and antibunched (i.e., sub-poissonian) photon statistics. Here we show experimentally that the spatially resolved autocorrelation function $g^{(2)}(0)$ of a pair of ions varies with the observation direction and the inter-ion distance and can take values below, above and equal to 1. Thus, we observe interference fringes of the intensity correlation function $g^{(2)}(0)$ phase shifted by $\pi$ from the electric field correlation interference stripes $G^{(1)}(0)$. 

In this letter we start working out the theory, using the Dicke model and outline the quantum description of emission from a pair of indistinguishable emitters, and deduce an analytic expression of the spatial dependence of $g^{(2)}(\boldsymbol{r},0)$. Then we describe the experimental apparatus including the movable Hanbury Brown and Twiss detector, and discuss the measured autocorrelation function. Finally, we explain the observed minimal and the observed maximal value of $g^{(2)}(0)$ from independently determined experimental parameters and sketch future application and research goals.  

\begin{figure*}
\includegraphics[width=\textwidth]{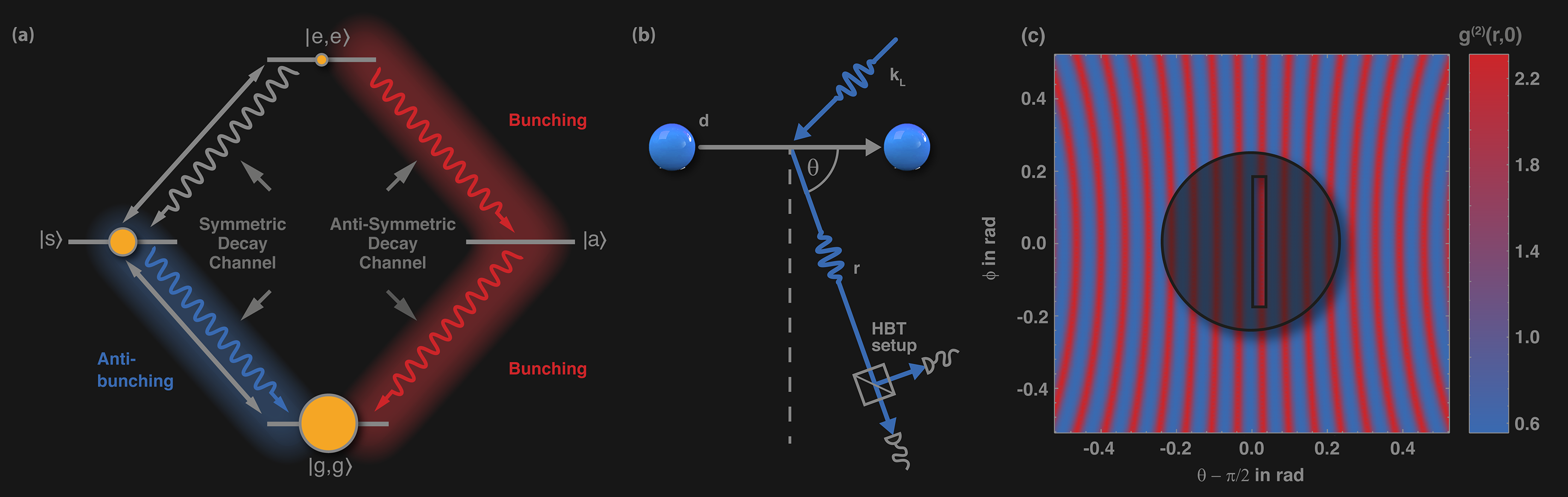}
\caption{(a) Dicke model of two identical two-level atoms: Out of the four states $\ket{g,g}, \ket{s}=(\ket{e,g}+\ket{g,e})/\sqrt{2}, \ket{a}=(\ket{e,g}-\ket{g,e})/\sqrt{2}, \ket{e,e}$, the states $\ket{g,g}$, $\ket{s}$ and $\ket{e,e}$ are symmetric and the state $\ket{a}$ is anti-symmetric. Laser light (gray double arrows) couples only the symmetric Dicke states and transfers population (indicated by the size of yellow circles) between $\ket{g,g}\leftrightarrow\ket{s}$ and $\ket{s}\leftrightarrow\ket{e,e}$, whereas spontaneous emission of a photon (shown as curly arrows) connects symmetric states (along symmetric decay channels) as well as anti-symmetric states (along anti-symmetric decay channels). (b) Due to the trapping potential and the mutual Coulomb repulsion, two ions are fixed in space at a distance $\boldsymbol{d}$. For the measurement of $g^{(2)}(\boldsymbol{r}, \tau)$ the two-ion crystal is illuminated with a laser propagating along $\boldsymbol{k}_L$. Scattered photons are recorded in the far field by a spatial-resolving Hanbury Brown Twiss setup under angle $\theta$. (c) Expected spatial modulation of $g^{(2)}(\boldsymbol{r}, \tau)$: regions with photon bunching $g^{(2)}(\boldsymbol{r}, \tau) > 1$ (red) alternate with regions with photon antibunching $g^{(2)}(\boldsymbol{r}, \tau)<1$ (blue). As the two-ion crystal is oriented in the plane with azimuthal angle $\phi=0$ we select the regions by a rectangular slit (rectangle). Averaging over the full solid angle of the light collection system (circle) washes out the spatial variation of the $g^{(2)}(\boldsymbol{r}, \tau)$.
}
\label{Fig1}
\end{figure*}

{\it Theory:} A pair of two-level identical atoms is advantageously described in the Dicke model, with symmetric and anti-symmetric combinations of the ground states $\ket{g}_i$ and excited states $\ket{e}_i$ of atoms, $i = 1,2$. For an interatomic separation $d \gg \lambda$ and negligible interactions Dicke states read \cite{skornia2001nonclassical,schon2001analysis} (see Fig. \ref{Fig1}a)
\begin{equation*}
\begin{aligned}
&\ket{e,e},\quad \ket{g,g}\\
&\ket{s}=\frac{1}{\sqrt{2}}\left(\ket{e,g}+\ket{g,e}\right)\\
&\ket{a}=\frac{1}{\sqrt{2}}\left(\ket{e,g}-\ket{g,e}\right)
\end{aligned}
\end{equation*}

Due to destructive interference, the dipole interaction with the laser vanishes for the anti-symmetric Dicke state. Thus, population is transferred by the laser interaction only between states $\ket{g,g}\leftrightarrow\ket{s}$ and $\ket{s}\leftrightarrow\ket{e,e}$. By contrast, spontaneous emission jump operators connect both, the symmetric as well as the anti-symmetric Dicke states (see Fig. \ref{Fig1}a). The transition amplitude depends on the position of observation. This can be seen when considering the photon detection operator \cite{skornia2001nonclassical}
\begin{equation}
\hat{D}=\hat{\sigma}_1+e^{i\delta (\boldsymbol{r})}\hat{\sigma}_2,
\label{eq:DecayOperator}
\end{equation}

expressing by use of the individual atomic lowering operators $\hat{\sigma}_1=\ket{g}_1\bra{e} (\hat{\sigma}_2=\ket{g}_2\bra{e})$ for atom 1 (atom 2) the two options that either the first or the second atom emits a photon via spontaneous decay. Since it is assumed that the photon detection occurs in the far field, the detector cannot distinguish between these two possibilities, hence both options are considered in Eq. \ref{eq:DecayOperator}. The operator $\hat{D}$ takes further into account the difference in optical phase $\delta(\boldsymbol{r})=(\boldsymbol{k}_L-k\hat{\boldsymbol{r}})\cdot\boldsymbol{d}=\boldsymbol{k}_L\cdot\boldsymbol{d}-kd\cos\theta$ between the two possibilities, accumulated by a photon when scattered by atom 1 with respect to a photon scattered by atom 2. Here, $\boldsymbol{k}_L$ is the wavevector of the incoming laser beam, $k=2\pi/\lambda\approx k_L$ and $\lambda$ are the wavenumber and wavelength of the spontaneously emitted photon, respectively, $\boldsymbol{d}$ the distance vector between the two atoms, $\hat{\boldsymbol{r}}$ the unit vector pointing towards the detector at $\boldsymbol{r}$, and $\theta$ the polar angle with respect to the ion crystal axis (see Fig. \ref{Fig1}b).

If the detector is placed at a position $\delta_s$ such that $\delta(\boldsymbol{r})$ is an even integer multiple of $\pi$ the detection operator $\hat{D}$ is symmetric and thus connects only the symmetric Dicke states. For low laser intensities most of the atomic population resides in the ground state $\ket{g,g}$, whereas the state $\ket{s}$ is only little populated and the state $\ket{e,e}$ is hardly occupied at all. Hence, a spontaneously emitted photon will most likely stem from state $\ket{s}$, going along with the transition $\ket{s}\rightarrow\ket{g,g}$, see Fig.~\ref{Fig1}a. The dynamics of the laser-driven two-atom system within the four levels of the Dicke model thus takes place essentially between the two levels $\ket{g,g}\leftrightarrow\ket{s}$, resulting in {\it antibunching statistics} like of a light-emitting single particle. However, if the detector is located at a position $\delta_a$ such that $\delta(\boldsymbol{r})$ is an odd integer multiple of $\pi$, the detection operator $\hat{D}$ becomes anti-symmetric. Now, photon detection originates exclusively from the decay channel $\ket{e,e}\rightarrow\ket{a}$, followed by $\ket{a}\rightarrow\ket{g,g}$ thus resulting in {\it bunching of photons} in pairs. Single photon events are excluded, because the anti-symmetric state $\ket{a}$ can not be populated by the driving laser field from the ground state $\ket{g,g}$ due to quantum interference, see Fig.~\ref{Fig1}a. In conclusion, the emission of a two-ion crystal features a remarkable change in photon statistics. Just by moving the detector from $\delta_s$ to $\delta_a$, the autocorrelation function varies between antibunching and bunching, while all parameters of the source are kept unchanged.

The detailed calculation shows that the spatial dependence of the autocorrelation function $g^{(2)}(\boldsymbol{r},\tau)$ at $\tau=0$ for two immobile laser driven two-level atoms is given by \cite{skornia2001nonclassical}
\begin{equation}
g^{(2)}(\boldsymbol{r},0)=\frac{(1+s)^2}{\left(1+s+\cos\delta(\boldsymbol{r})\right)^2},
\label{eq:g2Theo}
\end{equation}

where $s$ denotes the laser saturation of the transition $\ket{g}_i\leftrightarrow\ket{e}_i$ (assumed to be identical for both atoms $i = 1,2$). The function $g^{(2)}(\boldsymbol{r},0)$ displays a spatial modulation with a minimal value $<1$ at $\delta_s$ and a maximal value $>1$ at $\delta_a$. The modulation of $g^{(2)}(\boldsymbol{r},0)$ reveals that the light emitted by a two-ion crystal displays a fundamentally different behavior than the light emitted by a single atom. While the latter exhibits solely antibunching, independent of the angle of observation, the former varies in space, undergoing a continuous crossover from antibunching, to laser-like emission of uncorrelated photons, to photon bunching. 

The detection of one spontaneously emitted photon is heralding the two-ion crystal in one particular entangled state. For example, recording a photon along the direction for the anti-symmetric decay channel projects the system into the maximally entangled anti-symmetric state $\ket{a}$. We are selecting this projective measurement in free-space. For the case of qubits in waveguide structures, the phenomenon has been studied theoretically \cite{zhang2019heralded}, because it is useful for generating entanglement of distant qubits, e.g. for quantum repeaters \cite{moehring2007entanglement,hofmann2012heralded,bernien2013heralded,slodivcka2013atom,delteil2016generation,stockill2017phase}.

\begin{figure}
\includegraphics[width=0.45\textwidth]{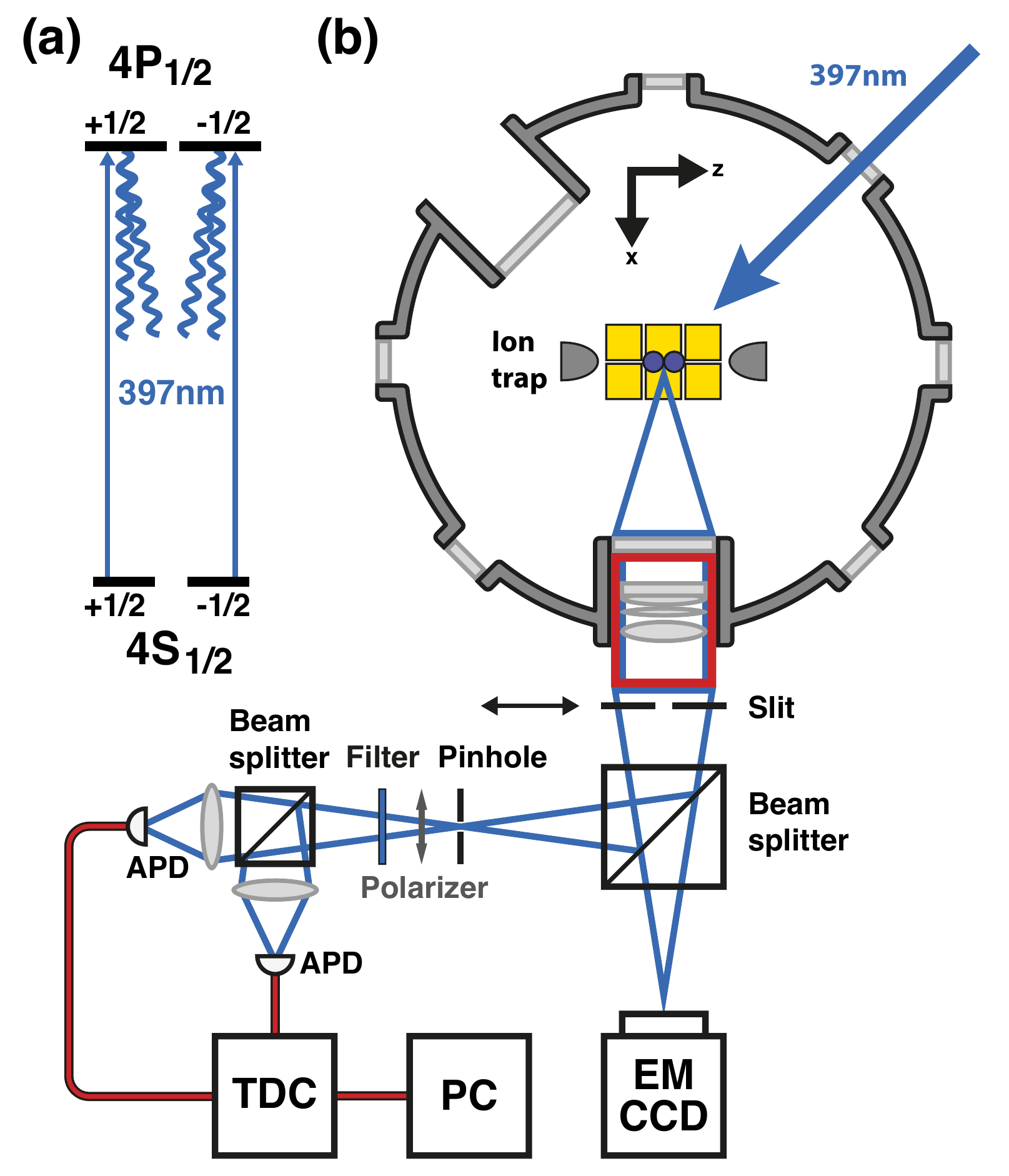}
\caption{
(a) Relevant level scheme of the $^{40}$Ca$^+$ $\text{S}_{1/2}\rightarrow\text{P}_{1/2}$ transition at $\lambda = 397\,$nm used in the experiment, including Zeeman-sublevels. (b) Experimental setup: Two ions, held in a segmented microtrap (yellow), are illuminated by a laser near $397\,$nm, forming a linear crystal along the $z$ axis of the trap. The fluorescence light is collected by an imaging objective. A movable slit aperture limits the observation to a small solid angle to allow for the discrimination of different observation directions. The light is focused into a Hanbury Brown and Twiss setup.\\
}
\label{Fig2}
\end{figure}

{\it Experiment:} In order to reveal the features of the autocorrelation function, we trap and laser cool two $^{40}$Ca$^+$ ions in a segmented linear Paul trap \cite{jacob2016transmission}.  The ions form a linear crystal along the $z$ axis of the trap (trap frequencies $\omega_{z,R1,R2}/2\pi= (0.760,1.275,1.568)\,$MHz) with interatomic distance $d= 6.7\,\mu$m. Doppler cooling is employed with a laser near $397\,$nm, red detuned by $2\pi\times 30\,$MHz with respect to the $\text{S}_{1/2}\rightarrow\text{P}_{1/2}$ transition. The orientation of its $\boldsymbol{k}_L$-vector allows for continuous cooling of all eigenmodes of the crystal, see Fig. \ref{Fig2}. Doppler cooling results in a residual motion of the ions of $\Delta x < 100\,$nm. A magnetic field of $0.62\,$mT along the $y$-direction, generated by three pairs of coils, defines the quantization axis. The fluorescence light emitted by the ions is collected with a $f/1.6$ objective at a working distance of $48.5\,$mm \cite{wolf2016visibility}. To resolve the spatial variation of $g^{(2)}(\boldsymbol{r}, \tau)$, a movable vertical $1\,$mm slit aperture behind the objective is introduced blocking all fluorescence light except for a small angle (see Fig. \ref{Fig1}c).

The vertical sheet of fluorescence light reflected by a beam splitter, is imaged onto a pinhole with diameter $200\,\mu$m to reduce residual background. Indistinguishability of the photon polarization is ensured by introducing a polarizing filter behind the pinhole, transmitting $\pi$-polarized light only. The light, after passing a bandpass filter at $400\,$nm, is fed into a Hanbury Brown and Twiss setup, consisting of a beam splitter and two fibre-coupled single-photon avalanche photo diodes (APD) with $85\,\%$ detection efficiency, dark count rate of $10\,$Hz and timing jitter of $1.6\,$ns. Time-tags for each photon detection event are recorded by a time-to-digital converter (TDC) and subsequently correlated via software on a PC. 

{\it Automatic data acquisition:} To prevent false data acquisition, a $90:10$ beam splitter behind the $1\,$mm slit aperture transmits $10\,\%$ of the scattered light to image the two-ion crystal on an electron-multiplying charge coupled device camera (EMCCD), see Fig. \ref{Fig2}. In this way the number of ions is constantly monitored using a fluorescence threshold technique. Upon ion loss, the measurement is automatically interrupted, and exactly two ions are reloaded into the trap. Only then, the data acquisition is continued. A $24/7$ operation is crucial for obtaining sufficient statistics. Within an acquisition time of ~$ 72\,$h, about $300$ correlation events per $2\,$ns bin were collected for each position over a temporal window of $600\,$ns. 

\begin{figure*}
\includegraphics[width=\textwidth]{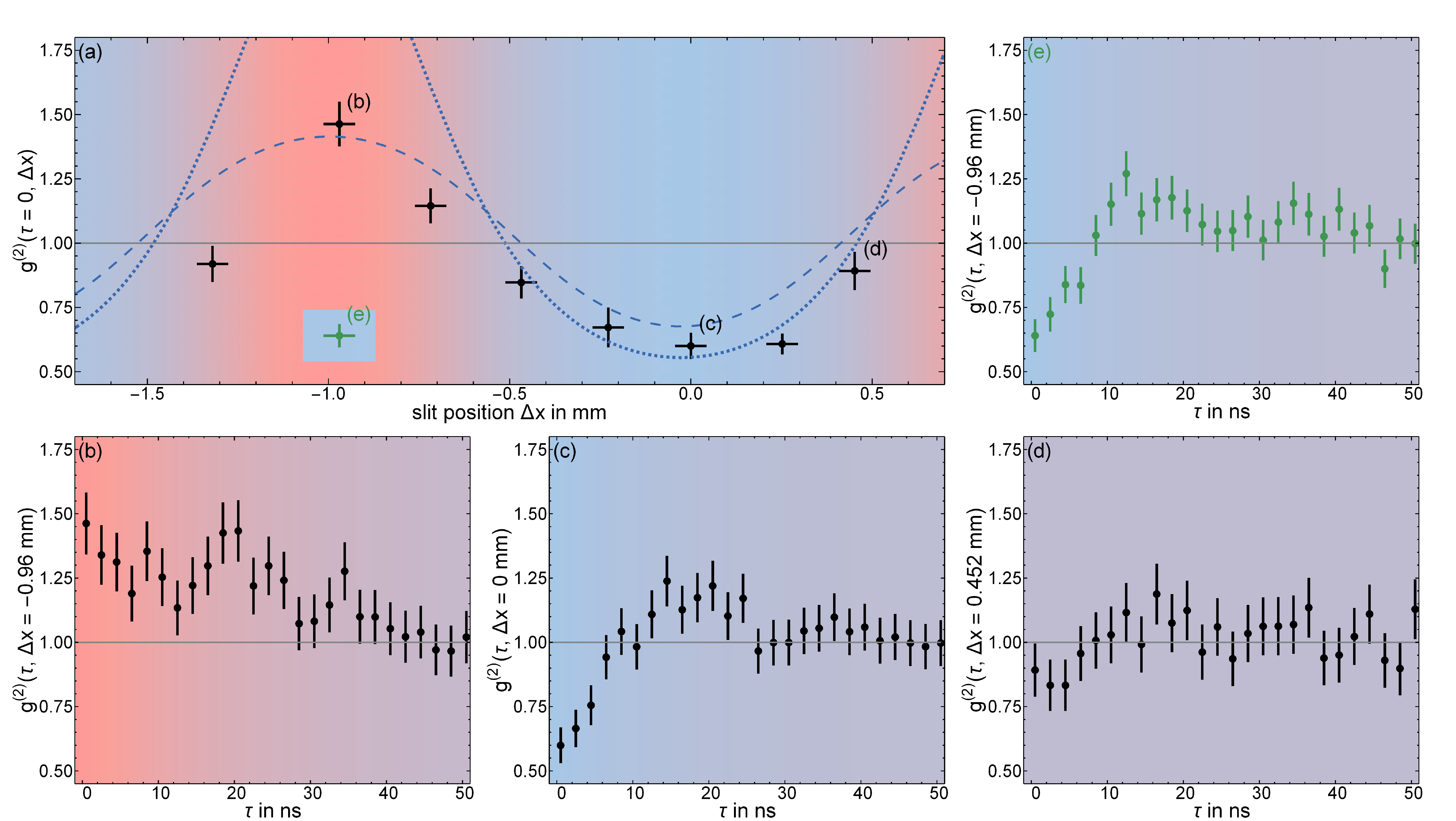}
\caption{
(a) Experimental results of the two-photon autocorrelation function $g^{(2)}(\boldsymbol{r}, 0)$ for eight different detector positions. The different regimes (bunching, laser-like, antibunching) can be clearly observed. (b) – (d): temporal dependence of $g^{(2)}(\boldsymbol{r}, \tau)$ for bunched (b), antibunched (c), and laser-like (d) photon statistics for a time window of $600\,$ns with $2\,$ns bin size. Taking into account the laser saturation of the $^{40}$Ca$^+$ $\text{S}_{1/2}\rightarrow\text{P}_{1/2}$ transition and the residual motion of the atoms after Doppler cooling, a spatial modulation of $g^{(2)}(\boldsymbol{r}, \tau)$as indicated by the blue dotted line is predicted. Considering additionally the multi-level structure of $^{40}$Ca$^+$, i.e., the possible decay into the metastable $3^2\text{D}_{3/2}$-level, as well as dark counts and the limited spatial and temporal resolution of the detection process, leads to a modified contrast $g^{(2)}(\boldsymbol{r}, 0)$  (dashed line). (e) By altering the inter-ion distance the observed $g^{(2)}(\boldsymbol{r}, 0)$-value changes from bunching to antibunching.
}
\label{Fig3}
\end{figure*}

{\it Results:} The arrangement allows for determining the spatial and temporal dependence of the autocorrelation function $g^{(2)}(\boldsymbol{r}, \tau)$. The latter was measured at eight different observation angles. Prior to each measurement point of the $g^{(2)}(\boldsymbol{r},\tau)$-signal, the position of the fringe pattern is checked by integrating the added signal from both APDs over one minute for different slit positions. In this way the $G^{(1)}$-signal is obtained, from which the signal periodicity is fitted. The slit position in Fig. \ref{Fig3} is given as an offset to the position of the center interference maximum from these $G^{(1)}$-measurements, with error bars obtained from the $G^{(1)}$-fit parameter standard errors.

For a slit position of $0\,$mm, corresponding to $\delta(\boldsymbol{r})=\delta_s$, we measure $g^{(2)}(\boldsymbol{r}, \tau) = 0.60(5)$, i.e. antibunched photon statistics. By contrast, for a slit position at $0.45\,$mm, corresponding to $\delta_u= 89(2)^\circ$, we observe uncorrelated photon arrival times with $g^{(2)}(\boldsymbol{r}, \tau)= 0.89(7)$. Finally, for a slit position at $-0.97\,$mm, corresponding to $\delta(\boldsymbol{r})=\delta_a$, we record $g^{(2)}(\boldsymbol{r}, \tau)=1.46(8)$, i.e. bunched photon statistics. Three examples of different {\it temporal} emission characteristics are plotted in Fig.~\ref{Fig3}b-d. Note, that we vary only the angle of observation while keeping all parameters such as laser power with $s=0.46(8)$, laser detuning $\Delta=2\pi\times 30(3)\,$MHz, trap frequency etc. unchanged.

An alternative way for tailoring the photon statistics of the two-ion crystal is a change of the inter-ion distance. This varies the optical phase difference $\delta(\boldsymbol{r})$, even while the observation angle remains unchanged. Reducing the axial trap frequency $\omega_z$ from $760\,$kHz to $718\,$kHz, the ion distance increases to $d=6.97\,\mu$m and $\delta$ by $0.96\,\pi$. Now, the two-photon autocorrelation function switches to antibunching, with $g^{(2)}= 0.64(4)$ (see green data point in Fig. \ref{Fig3}a and \ref{Fig3}e).

Our aim is to understand the experimental data from a complete model and independently determined parameters: The saturation parameter $s$ is measured via the autocorrelation of the light scattered by a single ion allowing for the observation of Rabi oscillations. From the frequency of g$^{(2)}$-oscillations and the measured detuning of the excitation laser, the saturation parameter determined to $s=0.46(8)$. The ion motion, after Doppler cooling with a phonon number in every mode of $n\approx10$ is leading to Debye-Waller-Factor $f_{DW}=0.50(5)$ is taken into account. The spatial periodicity of the $g^{(2)}$-signal is determined as $L=1.94(4)\,$mm, using the $G^{(1)}$-signal of a two-ion crystal. This model leads to the prediction shown in Fig. \ref{Fig3}, dashed line, with only the horizontal offset as a fit parameter. A maximum of $g_{max}^{(2)}(0)=2.31(27)$ is expected for bunching and $g_{min}^{(2)}(0)=0.55(03)$ for antibunching, respectively.

Including several mechanisms limiting the achievable $g^{(2)}$-contrast refines the model and improves the agreement with measured data. Eq. \ref{eq:g2Theo} is derived from a theory based on two-level atoms. Yet, $^{40}$Ca$^+$ ions have a more complex level structure. If the ions decay into the metastable $3\text{D}_{3/2}$-level (branching ratio $\sim$ 1:16 \cite{hettrich2015measurement}), only a single ion remains in the trap that may scatter photons at $397\,$nm, now showing perfect antibunching. Additionally, the slit has a finite width of $1\,$mm to collect a sufficient number of photons. This leads to an averaging of the $g^{(2)}$-signal over this spatial detection angle, further reducing the contrast. Also, the temporal uncertainty of $1.6\,$ns of the employed photo detectors averages any temporal behavior of the ions, governed by an exponential decay towards $g^{(2)}(\tau)=1$ with a time constant given by the excited state lifetime of $\tau=6.9\,$ns \cite{hettrich2015measurement}. Finally, the dark count rate of the detectors leads to false start and stop signals of the $g^{(2)}$-measurement giving rise to a constant offset of the $g^{(2)}$-signal. When summarized, the listed contrast limiting effects lead to $g^{(2)}_{max}(0)=1.41(12)$ for bunching and $g^{(2)}_{min}(0)=0.68(2)$ for antibunching, respectively. This model is shown in Fig. \ref{Fig3} (blue line)  together with its confidence interval (light blue region). Taking into account the eight measured values in Fig. \ref{Fig3}, we find good agreement with the expected spatial modulation of $g^{(2)}(\boldsymbol{r}, \tau)$, derived from the independent evaluation. 

In the future, we plan spatio-temporal two-photon correlation measurements with increased contrast allowing for the observation of superbunching \cite{bhatti2015superbunching}. We further anticipate impact of our method for quantum computing \cite{jiang2007distributed,luo2009protocols,duan2010colloquium} and quantum communication \cite{briegel1998quantum,duan2001long}, when distant processor units are connected via projective measurements \cite{moehring2007entanglement,hofmann2012heralded,bernien2013heralded,slodivcka2013atom,delteil2016generation,stockill2017phase}, possibly for larger ion crystals \cite{obvsil2019multipath}. The measured heralded quantum entanglement of states can also be converted to a tailored decay yielding long-lived entangled spin states. A designed photon statistics allows furthermore for improved imaging techniques, useful the life sciences \cite{dertinger2009fast,schwartz2013superresolution,classen2018analysis}. Measuring cross-correlations instead of autocorrelations may finally disclose the non-classicality of the photon correlations \cite{wiegner2010creating}, applicable for ultra-precise sensing, intensity interferometry \cite{basche1992photon,skornia2001nonclassical,schon2001analysis,richter1991interference} as well as structure analysis in crystallography \cite{classen2017incoherent}.

\begin{acknowledgments}
SR and JvZ acknowledgment support by the School of advanced optical technologies, Erlangen and and the International Max-Planck Research School, Physics of light, Erlangen.
\end{acknowledgments}

\bibliographystyle{apsrev4-1}
\bibliography{lit}
\onecolumngrid

\end{document}